\documentclass[conference]{IEEEtran}
\usepackage{epsf}
\usepackage{graphicx}
\usepackage{amsmath,bm}
\usepackage{amssymb}
\usepackage{epsfig,latexsym,amsmath,epsf,amssymb,amsfonts}
\usepackage{verbatim}
\usepackage{placeins}
\usepackage[hang]{subfigure}
\usepackage{epstopdf}
\usepackage{amsthm}
\usepackage{url}
\usepackage{algorithm}
\usepackage{algorithmic}
\newcommand{\erfc}{\mathrm{erfc}\,}
\usepackage{caption}
\usepackage{cite}

\begin{document}

\pagestyle{empty}

\title{Maximum Throughput Opportunistic Network Coding in Two-Way Relay Networks}

\author{\IEEEauthorblockN{Maha Zohdy, Tamer ElBatt, and Mohamed Nafie}
\IEEEauthorblockA{Wireless Intelligent Networks Center (WINC) \\
Nile University, Giza, Egypt\\
m.zohdy@nileu.edu.eg, telbatt@ieee.org, mnafie@nileuniversity.edu.eg}
}
\vspace{-0.2cm}
\maketitle

\thispagestyle{empty}

\begin{abstract}
In this paper, we study Two-Way Relaying $\left(\textbf{TWR}\right)$ networks well-known for its throughput merits. In particular, we study the fundamental throughput delay trade-off in two-way relaying  networks using opportunistic network coding $\left(\textbf{ONC}\right)$. We characterize the optimal \textbf{ONC} policy that maximizes the aggregate network throughput subject to an average packet delay constraint. Towards this objective, first, we consider a pair of nodes communicating through a common relay and develop a two dimensional Markov chain model capturing the buffers' length states at the two nodes. Second, we formulate an optimization problem for the delay-constrained optimal throughput. Exploiting the structure of the problem, it can be cast as a linear programming problem. 
 Third, we compare the optimal policy to two baseline schemes and show its merits with respect to throughput, average delay and power consumption. The numerical results reveal interesting insights. First, the optimal policy significantly outperforms the first baseline with respect to throughput, delay and power consumption. Moreover, it outperforms
the second baseline with respect to  the average delay and power consumption for asymmetrical traffic arrival rates.
\end{abstract}

\section{Introduction}
Triggered by the seminal work of Alshwede et al. \cite{networkinfromationflow}, network coding has received considerable attention from the community, initially, to improve wired networks capacity \cite{linearnetworkcoding}. Later, the merits of network coding prevailed in wireless networks due to the broadcast nature of the transmissions, opening up more opportunities for packet mixing \cite{informationexchange, codedrelaying, bat}.
Recently, there has been increasing interest from the community in studying network coding in cooperative relay networks. Cooperation in wireless networks \cite{usercooperation1},\cite{usercooperation2} constitutes one manifestation of spatial diversity which utilizes the broadcast nature of wireless transmissions to overcome the imperfections in wireless channels. TWR networks \cite{twowayrelaying} have recently emerged as one of the basic forms of cooperative networks. 

Network coding schemes in TWR networks can be divided into two generic schemes referred to as 3-step and 2-step schemes \cite{physicalnetworkcoding}. In 3-step schemes, each source node is allowed to transmit one packet to the relay in an exclusive time slot. In the third time slot, the relay broadcasts a bit-wise XOR-ed packet to both nodes to complete the two-way relaying process. On the other hand, 2-step schemes allow both source nodes to transmit simultaneously to the relay in one time slot. In the second time slot, the relay broadcasts 
the XOR-ed packet to both of them. In \cite{physicalnetworkcoding}, 
the authors investigated and characterized conditions for maximizing the two-way rate for a number of 3-step and 2-step schemes. In addition, the work in \cite{networkcodingfortwoway} characterized the achievable rate regions for 3-step network coding in TWR networks. In \cite{opportunistic}, the concept of opportunistic network coding was first introduced to minimize the delay encountered by packets waiting at the relay node to be encoded. In \cite{aylineyener , aylinecostdelay} the energy-delay trade-off was analyzed for conventional network coding. 
Most of the aforementioned works assume the presence of two buffers at the relay node in order to store packets from both sources to combine before relaying, using network coding. However, this setting leaves the multiple access channel (MAC) between the source nodes and the relay inefficiently utilized due to the stochastic nature of the packets arrivals at the source nodes. This, in turn, leaves no room for packet combining on the MAC channel resulting in throughput degradation. This work is motivated by the key observation that moving the buffers from the relay node to the source nodes would create more combining opportunities. This, in turn, results in ``additional'' throughput attained by other nodes in the network, as we show later in this paper. 
\begin{figure}[tb]
	\centering
	\includegraphics[width=2.5in,height=2in]{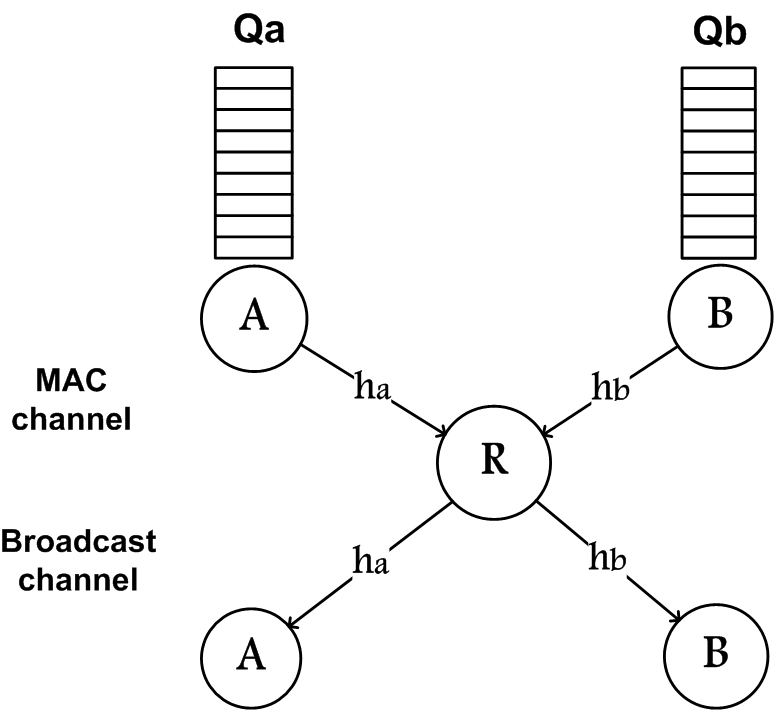}
	\caption{System Model}
	\label{fig:SystemModel}
	\vspace{-0.5cm}
\end{figure}

Our main contribution in this paper is three-fold. First, in contrast to \cite{opportunistic,aylineyener}, we move the buffering storage from the relay node to the source nodes which allows us to enhance the throughput of the MAC channel, and hence, the entire TWR network throughput. 
 Second, we characterize the optimal ONC policy, depending on the buffers' states and the packet arrival rates at the source nodes, that maximizes the total network throughput (including the aforementioned additional throughput) subject to average packet delay constraints. Third, we investigate the average transmission power consumed at the source nodes for the proposed policy. 
%
Under the proposed setting for packet buffers at the source nodes, we show that the TWR network can sustain the same throughput achieved by conventional network coding, yet, at a lower average delay. Moreover, the MAC channel can be utilized more efficiently to communicate the traffic of other nodes in the network, giving rise to ``additional'' throughput. 
Towards solving the aforementioned problem, we first construct a two-dimensional Markov chain capturing the buffers' states at the two source nodes at hand. Afterwards, we formulate and solve an optimization problem that maximizes the total average throughput of the network under a constraint on the average end-to-end packet delay. Finally, given the established optimal TWR transmission policy, we characterize the minimum average transmission power consumed by the source nodes under this policy and compare it to the baseline policies. 

The rest of the paper is organized as follows. In Section II, the system model and underlying assumptions are introduced. In Section III, the opportunistic TWR transmission scheme is introduced. Afterwards, the constrained optimization problem is formulated and solved efficiently to characterize the maximum average throughput subject to average delay constraints,in Section IV. Also, the average transmission power for the established optimal policy is characterized in the same section. The numerical results are presented and discussed in Section V for a number of scenarios. Finally, our conclusions are summarized in Section VI.

\section{System Model}
We consider a TWR network consisting of two source nodes $A$ and $B$ communicating through a relay node $R$, as shown in Fig. \ref{fig:SystemModel}. There is no direct link between node 
$A$ and node $B$. Unlike prior work, each source node is assumed to have a limited size buffer; denoted $Q_{a}$ of maximum length $N_{a}$ at node $A$ and $Q_{b}$ of maximum length $N_{b}$ at node $B$. While the relay node, $R$, has no queuing capability. Thus, in essence, one of the byproducts of this work is to shed light and quantify the benefits of a simple design change, that is, moving the packet buffers from the relay node (as in \cite{aylineyener,opportunistic} and many others) to the source nodes. As will be shown later, this simple design trick yields profound throughput gains attributed to a more efficient use of the MAC channel of TWR. In addition, we build upon this new system to characterize the optimal transmission policy at the source nodes to yield further performance gains beyond the state-of-the-art. 

We assume a time slotted system where each packet transmission fits exactly in one time slot. In addition, the two buffers' backlogs are assumed to be available at the relay node which is assumed to take the transmission decision every time slot. The TWR communication process between nodes $A$ and $B$ is divided into two steps \cite{bat};
\begin{itemize}
	\item Multiple access $($MAC$)$ step: nodes $A$ and$/$or $B$ transmit two$/$one packet(s) to the relay in one time slot.
	\item Broadcast $($BC$)$ step: the relay $R$ transmits the information back to the nodes in the subsequent slot. 
\end{itemize}

The packet arrival processes at source nodes $A$ and $B$ are independent and each follows an i.i.d. Bernoulli process with arrival rates $\lambda_{a}$ and $\lambda_{b}$ packets$/$slot, respectively, where $0\leq\lambda_{a},\lambda_{b}\leq1$. The channel between node $A$ (or $B$) and relay $R$ is assumed to suffer Rayleigh fading with scale parameter $s_{a}$ (or $s_{b}$). Let $h_{ar}$ denote the channel gain from node $A$ to node $R$. Similarly, the channel gains from node $B$ to node $R$, from node $R$ to node $A$ and from node $R$ to node $B$ are denoted by $h_{br}$, $h_{ra}$ and $h_{rb}$, respectively. The channels are assumed to be reciprocal, i.e. $h_{ar}$=$h_{ra}$=$h_{a}$ and $h_{br}$=$h_{rb}$=$h_{b}$ as shown in Fig. \ref{fig:SystemModel}. The channel state information (CSI) is assumed to be fully known at the source nodes as well as the relay node. All nodes are assumed to be half-duplex, that is, 
a node can receive an arriving packet at the beginning of time slot, if any, and hold it in its queue, but cannot transmit a received packet within the same time slot.
Let $P_{a}$ and $P_{b}$ denote the transmission power of nodes $A$ and $B$, respectively.
We assume that source nodes employ nested Lattice coding \cite{Lattice_coding} for simultaneous transmissions, such that in the event that two packets are received simultaneously at the relay node $R$, $P_{a}$ and $P_{b}$ are chosen so that they can be readily decoded. 
\section{Proposed Opportunistic network coding Two-Way Relaying Scheme}
In traditional network coding, e.g., \cite{aylineyener,opportunistic}, the arriving packets are transmitted by source nodes $A\left(or B\right)$ to node $R$, instantly, i.e. upon their arrival. In such setting, the packet buffers are assumed to reside only at the relay node, where a packet from one source always waits to be combined with a packet from the other source once it becomes available. However, this system setting has a fundamental limitation which causes the MAC channel to be busy at random time slots, governed by the packets arrival processes at nodes $A$ and $B$, transmitting individual ``uncombined'' packets. 
This, in turn, presents a major source of inefficiency in the MAC channel which we successfully remedy in this paper by using 2-step rather than the 3-step network coding approach.

The proposed scheme makes more efficient use of the MAC channel, since the buffers are assumed to be at the source nodes (instead of the relay node). This, in turn, makes it possible to control the time slots in which the MAC channel is occupied. 
%
This work quantifies the performance gains attributed to the optimal TWR transmission scheme under the new setting, in the sense of maximizing the total network throughput under an average delay constraint. We also characterize the average transmission power consumed by the optimal policy compared to two baseline schemes, including traditional network coding \cite{aylineyener}.

\subsection{Queuing-theoretic Model}
In order to analyze the performance of the proposed scheme, the buffers at nodes $A$ and $B$ are modeled as a two-dimensional Markov chain with steady state probabilities $\pi_{ij}$ where $i$ and $j$ represent the number of packets queued at nodes $A$ and $B$, respectively. 
The states of the Markov chain are denoted by $S(i,j)$, where $i=0,1,2,.....N_{a}$ and $j=0,1,2,......N_{b}$. The proposed randomized policy, assumed to be carried out by the relay node, takes state-dependent transmission decisions. 
Define $g^{k}_{ij}$, $k$ = 1,2,3,4, to denote the transmission probabilities under the 
four cases of interest: 
%
%
\begin{enumerate}
	\item Only node $A$ transmits one packet: with probability $g^{1}_{ij}$.
	\item Only node $B$ transmits one packet: with probability $g^{2}_{ij}$.
	\item Nodes $A$ and $B$ transmit two packets simultaneously using Lattice coding with probability $g^{3}_{ij}$.
	\item Neither node $A$ or $B$ transmit with probability $g^{4}_{ij}$.
\end{enumerate}
with the constraint that these probabilities should sum to one, that is, for each state $S(i,j)$,
\begin{equation}
	\sum_{k=1}^{4} g^{k}_{ij}=1
\end{equation}
Next, we characterize the probabilities of packets arrivals at the two source nodes, $A$ and $B$, in an arbitrary time slot. This gives rise to one of four cases, namely one packet arrival at each source, only one packet arrival at node $A$, one packet arrival at node $B$ and, finally, no packet arrivals. Hence, we define $f_i$, i = 1,2,3,4, to denote these four events, in order:
\begin{align}
f_{1}&= \lambda_{a}\lambda_{b}\\
f_{2}&= \lambda_{a}\left(1-\lambda_{b}\right)\\
f_{3}&= \left(1-\lambda_{a}\right)\lambda_{b}\\
f_{4}&= \left(1-\lambda_{a}\right)\left(1-\lambda_{b}\right)
\end{align}
Next, we characterize the state transition probabilities of the Markov chain at hand using the characterized arrival rates at $Q_{a}$ and $Q_{b}$ in (2)-(5), along with the transmission probabilities $g_{ij}^{k}$ as follows.

First, if both buffers are empty (i.e. the origin state), it is straightforward to notice that we have four possible transitions,
\begin{align}
	P\left(S\left(0,0\right) | S\left(0,0\right)\right) &=f_{4}\\
	P\left(S\left(0,1\right) | S\left(0,0\right)\right) &=f_{3}\\
	P\left(S\left(1,0\right) | S\left(0,0\right)\right) &=f_{2}\\
	P\left(S\left(1,1\right) | S\left(0,0\right)\right) &=f_{1}
\end{align}
Second, if only $Q_a$ is non-empty which corresponds to states on the horizontal axis, $S\left(i,0\right)$, $i=1,2,3,........N_{a}$,
\begin{align}
	&P\left(S\left(i,0\right) | S\left(i,0\right)\right) =f_{2} g_{i0}^{1} + f_{4} g_{i0}^{4} \\
	&P\left(S\left(i+1,0\right) | S\left(i,0\right)\right) =f_{2}g_{i0}^4\\
	&P\left(S\left(i-1,0\right) | S\left(i,0\right)\right) =f_{4}g_{i0}^1\\
	&P\left(S\left(i,1\right) | S\left(i,0\right)\right) =f_{1} g_{i0}^{1} + f_{3} g_{i0}^{4}\\
	&P\left(S\left(i+1,1\right) | S\left(i,0\right)\right) =f_{1}g_{i0}^4\\
	&P\left(S\left(i-1,1\right) | S\left(i,0\right)\right) =f_{3}g_{i0}^1
\end{align}
As for the vertical axis states, $S\left(0,j\right)$, $j=1,2,3,........N_{b}$, the state 
transition probabilities can be derived along the same lines of (10)-(15).
Finally, we consider the general case where both node $A$ and $B$ have non-empty buffers, i.e. the interior of the state space, 
$S\left(i,j\right)$ for $i=1,2,3,........N_{a}$,$j=1,2,3,........N_{b}$,
\begin{align}
	&P\left(S\left(i,j\right) | S\left(i,j\right)\right) =f_{4} g_{ij}^{4} + f_{1} g_{ij}^{3} + f_{3} g_{ij}^{2} + f_{2} g_{ij}^{1} \\
	&P\left(S\left(i-1,j-1\right) | S\left(i,j\right)\right) =f_{4}g_{ij}^{3}\\
	&P\left(S\left(i+1,j+1\right) | S\left(i,j\right)\right) =f_{1}g_{ij}^{4}
	\end{align}
	\begin{align}
	&P\left(S\left(i,j-1\right) | S\left(i,j\right)\right) =f_{2} g_{ij}^{3} + f_{4} g_{ij}^{2}\\
	&P\left(S\left(i,j+1\right) | S\left(i,j\right)\right) =f_{1} g_{ij}^{1} + f_{3} g_{ij}^{4}\\
	&P\left(S\left(i+1,j\right) | S\left(i,j\right)\right) =f_{1} g_{ij}^{2} + f_{2} g_{ij}^{4}\\
	&P\left(S\left(i-1,j\right) | S\left(i,j\right)\right) =f_{3} g_{ij}^{3} + f_{4} g_{ij}^{1}\\
	&P\left(S\left(i-1,j+1\right) | S\left(i,j\right)\right) =f_{3}g_{ij}^{1}\\
	&P\left(S\left(i+1,j-1\right) | S\left(i,j\right)\right) =f_{2}g_{ij}^{2}
\end{align}
Given the introduced Markov chain model, we formulate and solve the target optimization problem in the next section, aiming at maximizing the total network throughput under an average packet delay constraint.

\section{Optimal Throughput Policy under Delay Constraints}
\subsection{Problem Formulation}
In order to demonstrate the throughput gains on the MAC channel in our setting, which contributes to enhancing the total TWR throughput, we assume the presence of another source-destination pair denoted by $CD$, respectively. The pair $CD$ is assumed to utilize the MAC channel idle slots, i.e. not used by pair $AB$. These nodes may be readily thought of as low priority opportunistic users who are continuously on the look for spectrum holes (unused slots in our context), e.g. cognitive radio users with perfect sensing capability. 

Now let $\mu_{1}$ denote the throughput of nodes $A$ and $B$ while $\mu_{2}$ denotes that of nodes $C$ and $D$,
\begin{gather}
\mu_{1} = \sum_{i,j} \pi_{ij}\ast\left(\left(g_{ij}^{1} + g_{ij}^{2}\right) + 2\ast g_{ij}^{3} \right)\\
\mu_{2} = \sum_{i,j} \pi_{ij}\ast g_{ij}^{4}\\
\mu_{tot} = \mu_{1} + \mu_{2}
\end{gather}
We assume buffers $Q_{a}$ and $Q_{b}$ to be non-lossy thus (28) will be always satisfied.
\vspace{-0.1cm}
\begin{equation}
	\mu_{1} = \lambda_{a} + \lambda_{b}
\end{equation}
Thus by moving the buffers from the relay node to the source nodes $A$ and $B$, we can control the time slots in which the MAC  channel is empty along with maintaining (28). 

It is evident now from the above discussion that maximizing the total network throughput is equivalent to maximizing the number of empty slots on the MAC channel, represented by $\pi_{ij}g_{ij}^4$. This, in turn, gives rise to an instance of the fundamental throughput-delay trade-off which recurs in different problem contexts in wireless communications and networking. 
Motivated by this trade-off, we formulate a constrained optimization problem for maximizing the number of empty slots (equivalent to total throughput as shown in the previous discussion) subject to a 
constraint on the average packet delay, as follows:
\begin{align}
\textbf{P1} \hspace{1cm} &\max_{g_{ij}^{k}} \hspace{0.2cm} \sum_{i,j}\pi_{ij} g_{ij}^{4}\\ 
s.t. \hspace{1cm}&\frac{1}{\lambda_{a} + \lambda_{b}} \sum_{i,j}\pi_{ij} \left(i + j\right) \leq D_{max}
\end{align}
\begin{align}
\label{const_2}
& \boldsymbol\pi \textbf{P}  = \boldsymbol\pi ,\hspace{0.5cm}  \sum_{i,j}\pi_{ij} = 1\\
\label{eq_35}
&	g_{0j}^{1} = g_{i0}^{2} = g_{0j}^{3} = g_{i0}^{3} = 0\\
\label{eq_36}
&	g_{N_{a}j}^{3} = g_{iN_{b}}^{3} = 1
\end{align}
\noindent where $D_{max}$ is the average packet delay constraint and $\textbf{P}$ is the state transition probability matrix of the system Markov chain described in Section III. It should be 
noted here that the objective function in {\bf P1} is the average number of empty slots, characterized by the no transmission probability $g^4_{ij}$ introduced earlier. The first constraint is the average packet delay constraint obtained using Little's law. The constraints in (\ref{const_2}) are the balance equation and probability normalization condition for the Markov chain model, respectively.
Then, in order to govern the physics of the system, the probabilities in (\ref{eq_35}) are set to zero indicating no transmissions are possible from empty buffers. Also, to avoid any packet loss at the buffers, a source node always transmits a packet from a full buffer with probability one as in (\ref{eq_36}).
\subsection{Problem Complexity and Solution Approach}
In order to solve the above problem we need to obtain the steady state probabilities
of the Markov chain governing the system dynamics, 
$\pi_{ij}$. Motivated by the sheer complexity of getting a closed form expression for the system Markov chain to be plugged in {\bf P1}, we resort to a change of variables similar to \cite{ulukus} in an attempt to linearize the problem and, hence, significantly simplify the solution.

To this end, we introduce the intermediate variables $x_{ij}^{k}$ where $x_{ij}^{k}=\pi_{ij} g_{ij}^{k}$ for $i=1,2,3,........N_{a}$,
$j=1,2,3,........N_{b}$ and $k=1,2,3,4$. Thus, we can solve the transformed problem for $x_{ij}^k$ and then map it back to the transmission probabilities, $g_{ij}^k$, using (\ref{transformeq}).
\begin{equation}
\label{transformeq}
	g_{ij}^{k} = \frac{x_{ij}^{k}}{\sum_{k=1}^{4} x^{k}_{ij}}
\end{equation}
\noindent Accordingly, problem $\textbf{P1}$ can be transformed to the equivalent problem $\textbf{P2}$ as follows,
\begin{align}
	\textbf{P2} \hspace{1cm} &\max_{x_{ij}^{k}}   \sum_{i,j}x_{ij}^{4}\\ 
	s.t. \hspace{1cm}&\frac{1}{\lambda_{a} + \lambda_{b}} \sum_{i,j}\sum_{k=1}^{4} x_{ij}^{k} \left(i + j\right) \leq D_{max}\\
	& \textbf{Q x}  = 0 ,\hspace{0.5cm}  \sum_{i,j}\sum_{k=1}^{4}x_{ij}^{k} = 1\\
	\label{eq_28}
	&x_{i0}^{2} = x_{i0}^{3} = x_{0j}^{1} = x_{0j}^{3} = 0\\
	&x_{N_{a}j}^{1} = x_{N_{a}j}^{2} = x_{N_{a}j}^{4} = 0\\
\label{eq_31}
	&x_{iN_{b}}^{1} = x_{iN_{b}}^{2} = x_{iN_{b}}^{4} = 0
\end{align}
\noindent where \textbf{x} is the vector of the new optimization variables 
	$\left[ x_{00} , x_{01}^{1} , x_{01}^{2} , x_{01}^{3} , x_{01}^{4},......, x_{ij}^{1} , x_{ij}^{2} , x_{ij}^{3} , x_{ij}^{4} ,...............\right]$ and \textbf{Q} is the transition equations matrix in terms of the new variables $x_{ij}^{k}$. In addition, we have the boundary conditions in (\ref{eq_28})-(\ref{eq_31}) corresponding to the boundary conditions (\ref{eq_35}),(\ref{eq_36}) of problem $\textbf{P1}$.

\subsection{Average power consumption for the optimal transmission policy}
Given the optimal transmission policy characterized in Section IV-B to maximize the total network throughput, our prime goal in this subsection is to determine the average transmission power for each node depending on the available CSI. In order to sustain a constant transmission rate at the source nodes, the consumed power varies according to the varying channel gain, i.e. a node should increase its transmission power during poor channel conditions. Thus in order to avoid wasting high transmission power, we propose that  
a packet can only be transmitted from a node if the channel gain for the corresponding link is greater than or equal to a threshold $h_{th}$.
 For the assumed Rayleigh distributed channel gain realizations with unit scale parameter $s = 1$, the probability that the channel gain 
$h$ is greater than or equal to $h_{th}$ in a given time slot is given by:
\vspace{-0.3cm}
\begin{equation}
	P\left(h \geq h_{th}\right) = \exp\left(- \frac{h^{2}_{th}}{2}\right)
\end{equation}
Note that according to the theory of Lattice coding \cite{Lattice_coding}, it has been established that the source nodes can effectively achieve a rate of $\frac{1}{2} \log \left(0.5 + h* P\right)$ for a channel gain of value $h$ and transmission power of value $P$. Assuming normalized noise power at the source nodes $A$ and $B$, let $\alpha = \left(2^{2r} - 1\right)$ and $\beta = \left(2^{2r} - 0.5\right)$ denote the SNR in case of a single transmission or simultaneous transmissions, respectively. And $r$ represents the rate of transmission per time slot, where $r$ is $1$ when a packet is transmitted and zero when no packets are transmitted.

Recall that the optimal transmission probabilities $g_{ij}^{k*}$ resulting from $\textbf{P2}$ are state dependent. Thus, in order to minimize the average transmission power, we map these probabilities to a channel threshold for nodes $A$ and $B$ as follows:
\begin{align}
\label{power1}
	g^{1*}_{ij}&=\exp\left(- \frac{h^{2}_{th_{a,ij}}}{2}\right) \left(1 - \exp\left(- \frac{h^{2}_{th_{b,ij}}}{2}\right)\right)\\
	g^{2*}_{ij}&=\exp\left(- \frac{h^{2}_{th_{b,ij}}}{2}\right) \left(1 - \exp\left(- \frac{h^{2}_{th_{a,ij}}}{2}\right)\right)\\
	g^{3*}_{ij}&=\exp\left(- \frac{h^{2}_{th_{a,ij}}}{2}\right) \exp\left(-\frac{h^{2}_{th_{b,ij}}}{2}\right)\\
	\label{power4}
	g^{4*}_{ij}&=\left(1 - \exp\left(\!-\frac{h^{2}_{th_{a,ij}}}{2}\!\right)\!\right)\!\!\left(1 - \exp\left(- \frac{h^{2}_{th_{b,ij}}}{2}\!\right)\!\right)
\end{align}
Note that (\ref{power1})-(\ref{power4}), can be derived for a Rayleigh distributed channel, i.e. 
$g_{ij}^{1*}$=$P\left(h_{a,ij} \geq h_{th_{a,ij}}\right)$.$ P\left(h_{b,ij} \leq h_{th_{b,ij}}\right)$$\forall$ $i,j$. Averaging over all possible realizations of the Rayleigh distributed channel, the average power consumed at the source nodes in every time slot follows one of the three possibilities
i) If $h_{a,ij}\left(h_{b,ij}\right)\!\geq\!h_{th_{a(b)},ij}$ $\&$ $h_{b,ij}\left(h_{a,ij}\right)$$<$$h_{th_{b(a)},ij}$ then,
\begin{align}
&P_{a,ij}\!\left(P_{b,ij}\right)\!=\!\alpha\sqrt{\frac{\pi}{2}} \erfc\!\!\left(\!\frac{h_{th_{a(b)},ij}}{\sqrt{2}}\!\!\right)\!\!\left(\!\!1\!-\exp\left(\!\!-\frac{h_{th_{b(a)},ij}^{2}}{2}\!\right)\!\!\right)\\
&P_{b,ij}\left(P_{a,ij}\right)=0
\end{align}
\\
ii) If $h_{a,ij}\!\geq \!h_{th_{a},ij}$ $\&$ $h_{b,ij} \!\geq\! h_{th_{b},ij}$ then,
\begin{align}
&P_{a,ij}=\beta\sqrt{\frac{\pi}{2}}\erfc\!\!\left(\frac{h_{th_{a},ij}}{\sqrt{2}}\right)\!\!\left(\!\exp\left(-\frac{h_{th_{b},ij}^{2}}{2}\right)\!\right)\\
&P_{b,ij}=\beta\sqrt{\frac{\pi}{2}}\erfc\!\!\left(\frac{h_{th_{b},ij}}{\sqrt{2}}\right)\!\!\left(\!\exp\left(-\frac{h_{th_{a},ij}^{2}}{2}\right)\!\right)
\end{align}
\\	
iii) If $h_{a,ij}$$<$$h_{th_{a},ij}$ $\&$ $h_{b,ij}$$<$$h_{th_{b},ij}$ then,
\begin{equation}
P_{a,ij}=P_{b,ij}=0
\end{equation}

Using the above transmission policy, would significantly decrease the transmission power consumed by the source nodes as it aims to prevent wasting unnecessarily increased transmission power to overcome bad channel conditions. On the other hand, although the simultaneous transmissions would slightly increase the transmission power \cite{Lattice_coding}, the proposed policy results in lower average transmission power compared to conventional network coding as shown in section V. 

In the next section, we present our numerical results showing the merits of the optimal TWR transmission policy with respect to maximizing the total network throughput subject to average delay constraints in addition to the average transmission power consumption by the source nodes. 
\begin{figure}[t]
\centering
	\includegraphics[width=3.5in,height=2.5in]{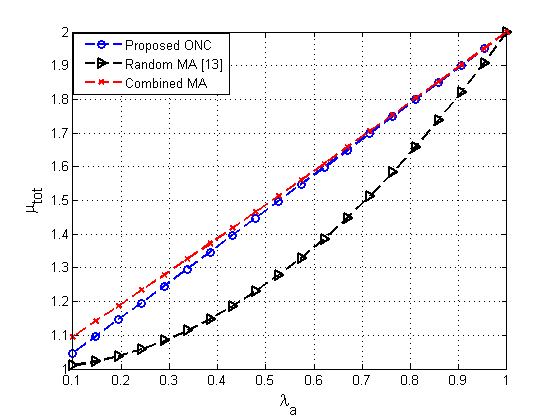}
	\caption{Total average throughput of nodes A, B, C and D, $\lambda_{a}$=$\lambda_{b}$.}
	\label{fig:TotalThroughput}
	\vspace{-0.4cm}
\end{figure}
\section{Numerical Results}
The numerical results presented in this section summarizes our major findings and lessons learned from this work. It is straightforward to establish that the formulated optimization problem $\textbf{P2}$ is a linear program that can be solved efficiently. In order to show the merits of the optimal policy, we will compare its performance to two baseline schemes, namely conventional network coding \cite{aylineyener,aylinecostdelay},  in terms of the total average throughput, average packet delay and the average transmission power consumption at the source nodes. 

As for the conventional network coding baselines, we assume having two systems, each consists of two source nodes $A$ and $B$ exchanging information through performing  conventional network coding at an intermediate relay node $R$. The first baseline scheme denoted by $Random$ MA is the same as that analyzed in \cite{aylineyener}, where node $R$ is assumed to have two finite length buffers. 
The source nodes transmit the arriving packets, whenever available, to node $R$, then every packet arriving at node $R$ waits to be combined with another packet from the opposite traffic direction. While in the second baseline denoted by $Combined$ MA, we assume having the same model but with the buffers now present in the source nodes instead of the relay thus the randomness in the MAC phase is no longer present. Note that in all of the three schemes, the broadcast channel operation is the same, such that the relay node broadcasts the received packet(s) after performing the Lattice coding operation if required.      
\begin{figure}[t]
\centering
	\includegraphics[width=3.5in,height=2.5in]{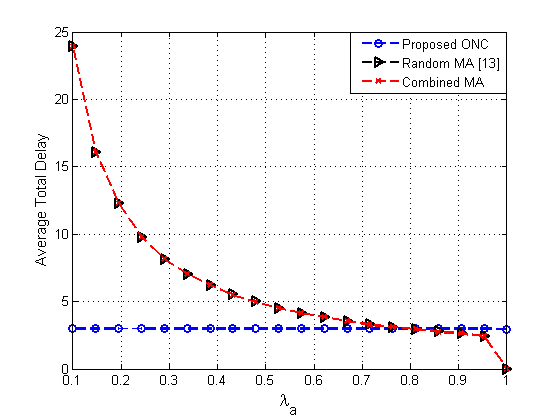}
	\caption{Total average delay at nodes A and B, $\lambda_{a}$=$\lambda_{b}$.}
	\label{fig:delay}
	\vspace{-0.5cm}
\end{figure}

We assume similar parameters in all the compared schemes as follows;  the buffers at both source nodes are of equal length where $N_{a} = N_{b} = 15$ packets,  packets arriving at nodes $A$ and $B$ follow an i.i.d Bernoulli process with equal average arrival rates $\lambda_{a}=\lambda_{b}$ packets$/$time slot.
 The total delay constraint applied to the proposed transmission scheme $D_{max} = 3$ time slots.  While in conventional network coding, the arriving packets from both directions are always queued awaiting a combining opportunity 
 whenever available which induces higher delay. Since all three transmission schemes studied are inherently non-lossy, the average sum throughput of the two source nodes $A$ and $B$ is equal to the sum average arrival rates.

In order to show the benefit attributed to increasing the average number of empty slots in the MAC channel by moving the buffers from the relay node to the source nodes, we will consider the presence of the additional nodes $C$ and $D$, with perfect sensing capabilities as described previously in section IV.  These nodes try to access the same channel only during these empty slots, to exchange some information packets of their own. For fair comparison, we assume that nodes $C$ and $D$ are also present in the other two baseline systems $Random$ MA and $Combined$ MA and operate only during empty slots, in a manner exactly similar to our proposed policy. 

Fig. \ref{fig:TotalThroughput} shows the total average throughput $\mu_{tot}$ of the overall system (four nodes: original nodes $A$ and $B$ along with the opportunistic users $C$ and $D$) making use of the spared empty slots in the given system. We assume that nodes $C$ and $D$ always have queued packets to transmit. It can be easily noticed that the total network throughput significantly increases when the buffers are present at the source nodes where the MAC channel can be efficiently used. The total throughput achieved from the proposed optimal scheme approaches this of the second baseline, $Combined$ MA. In this case, nodes $A$ and $B$ can transmit simultaneously, large portion of the time, which also increases the number of empty slots on the average for node $C$ and $D$. 

In the baseline systems using $Random$ MA and $Combined$ MA, the queued packets incur the same delay as they both use conventional network coding whether at the node $R$  or at nodes $A$ and $B$. And since the proposed transmission scheme is designed under a total delay constraint, Fig. \ref{fig:delay} shows that the previous throughput gain can be acquired under the required delay constraint. Thus our transmission scheme approaches the total throughput gain of system using $Combined$ MA, giving the maximum throughput gain, however with significantly lower delay.
\begin{figure}[t]
\centering
	\includegraphics[width=3.5in,height=2.5in]{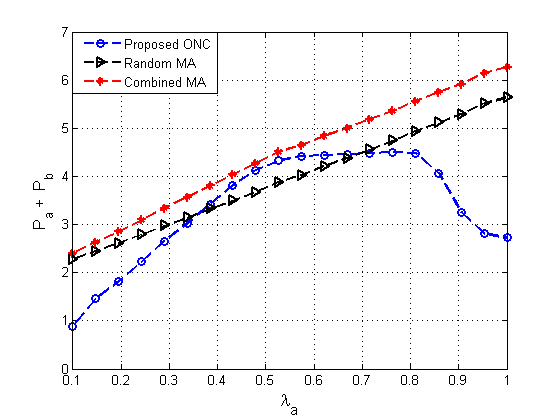}
	\caption{Average power of nodes A and B, $\lambda_{b}$=0.5.}
	\label{fig:power}
	\vspace{-0.5cm}
\end{figure}

The second part of the results is about the average power consumed at nodes $A$ and $B$ for asymmetrical arrival rates, $\lambda_{a}=0.5$ and $\lambda_{b} \in \left[0.1,1\right]$ packets/time slot. Given the transmission probabilities of the proposed scheme, the minimum average power consumption at nodes $A$ and $B$ is shown in Fig. \ref{fig:power} based on the mapping mentioned previously in section IV. Since the second baseline, $Combined$ MA, uses conventional network coding and the source nodes are aware of simultaneous transmission, the average power consumed is maximum with different arrival rates. On the other hand, the source nodes in $Random$ MA transmits their packets to the relay node upon arrival since the relay is the node responsible for combination. Thus nodes $A$
 and $B$ do not have to increase their transmission power in this case consuming lower power consumption. Finally, the proposed ONC scheme achieves the least power consumption levels for asymmetrical arrival rates approaching that of $Combined$ MA only at equal arrival rates, $\lambda_{a}=\lambda_{b}=0.5$ packets/time slot where the system tends to combine all the arriving packets before transmission.
\section{Conclusion}
In this paper, we studied the throughput delay trade-off in TWR networks using ONC. Specifically, we considered a pair of nodes, each having a finite length buffer, communicating through a common relay node. We characterized the optimal ONC transmission policy that maximizes the aggregate network throughput subject to an average packet delay constraint. First, we formulated an optimization problem for the delay-constrained optimal throughput. Then, we compared the performance of the proposed transmission policy to two other baseline schemes. The numerical results showed that the optimal policy outperforms the first baseline with respect to the average throughput, delay and power consumption. Also, it outperforms the second base line with respect to the average delay and power consumption for asymmetrical traffic rates. It is of interest as a future work, to investigate the general optimal transmission policy where the source nodes as well as the relay node both have queuing capabilities.

\bibliographystyle{IEEEtran}

\begin{thebibliography}{9}
\bibitem{networkinfromationflow}R. Ahlswede, N. Cai, S.-Y. R. Li, and R. W. Yeung, ``Network information
flow,'' in \textit{IEEE Trans. Inf. Theory}, vol. 46, no. 4, pp. 1204 - 1216, Jul 2000.

\bibitem{linearnetworkcoding}S. -Y. R. Li, R. W. Yeung and N. Cai, ``Linear network coding,'' in \textit{IEEE
Trans. Inform. Theory}, vol. 49, pp. 371 - 381, Feb 2003.


\bibitem{informationexchange}Y. Wu, P. A. Chou, and S.-Y. Kung, ``Information exchange in wireless
networks with network coding and physical-layer broadcast,'' in \textit{Microsoft
Research Technical Report MSR-TR-2004-78}, Aug 2004.

\bibitem{codedrelaying}P. Larsson, N. Johansson, and K.-E. Sunell, ``Coded bi-directional relaying,'' in \textit{ 5th Scandinavian Workshop on Ad Hoc Networks (ADHOC-05)}, Stockholm, Sweden, May 2005.

\bibitem{bat}P. Popovski and H. Yomo, ``Bi-directional amplification of throughput in a wireless multi-hop network,'' in \textit{IEEE 63rd Vehicular Technology Conference (VTC)}, Melbourne, Australia, May 2006.

\bibitem{usercooperation1}A. Sendonaris, E. Erkip, and B. Aazhang, ``User Cooperation Diversity- Part I:System Description,'' in \textit{IEEE transactions on communications}, vol. 51, no. 11, Nov. 2003

\bibitem{usercooperation2}A. Sendonaris, E. Erkip, and B. Aazhang, ``User Cooperation Diversity- Part II:Implementation aspects and system analysis,'' in \textit{IEEE transactions on communications}, vol. 51, no. 11, Nov. 2003

\bibitem{twowayrelaying}B. Rankov and A. Wittneben, ``Spectral efficient protocols for half-duplex
fading relay channels,'' in \textit{IEEE J. Select. Areas Commun.}, vol. 25, no. 2,
pp. 379-389, Feb. 2007.

\bibitem{physicalnetworkcoding}Petar Popovski and Hiroyuki Yomo, ``Physical Network Coding in Two-Way Wireless Relay Channels,''
in \textit{ Proc. IEEE ICC} ,Glasgow ,June,2007.

\bibitem{networkcodingfortwoway}Chun-Hung Liu and Feng Xue, ``Network Coding for Two-Way Relaying: Rate Region,
 Sum Rate and Opportunistic Scheduling,'' in \textit{Proc. IEEE ICC} ,Beijing  ,May 2008.

\bibitem{opportunistic}W.Chen, K.B. Letaief, Z.Cao, ``Opportunistic Network Coding for Wireless Networks,''
in \textit{Proc. IEEE ICC} ,Glasgow ,June 2007.

\bibitem{aylineyener}Xiang He and Aylin Yener, ``On the Energy-Delay Trade-off of a Two-Way
Relay Network,'' in \textit{Information Sciences and Systems}, 2008. CISS 2008. 42nd Annual Conference.

\bibitem{aylinecostdelay}E.N.Ciftcioglu, Y.E.Sagduyu, R.A. Berry, A.Yener, ``Cost-Delay Tradeoffs for Two-Way Relay Networks,''
in \textit{IEEE transactionson wireless communications}, vol. 10, no. 12, December 2011.

\bibitem{Lattice_coding}Makesh Pravin Wilson and Krishna Narayanan , ``Power Allocation Strategies and Lattice Based Coding schemes for
Bi-directional relaying,'' 
in \textit{in IEEE ISIT}, July 2009.


\bibitem{ulukus}Jing Yang and  Sennur Ulukus, ``Delay-Minimal Transmission for Average Power Constrained Multi-Access Communications,'' 
in \textit{IEEE transactions on wireless communications,} vol. 9, no. 9, Sep. 2010.


\end{thebibliography}

\end{document}